\documentstyle[epsfig]{mn-1.4}
\begin{document}

\def\ltsima{$\; \buildrel < \over \sim \;$}
\def\simlt{\lower.5ex\hbox{\ltsima}}
\def\gtsima{$\; \buildrel > \over \sim \;$}
\def\simgt{\lower.5ex\hbox{\gtsima}}
\def\ls{{_<\atop^{\sim}}}
\def\lax{{_<\atop^{\sim}}}
\def\gs{{_>\atop^{\sim}}}
\def\gax{{_>\atop^{\sim}}}
\def\cgs{ ${\rm erg~cm}^{-2}~{\rm s}^{-1}$ } 
  
\title[The HELLAS survey]
{The BeppoSAX High Energy Large Area Survey HELLAS, III: 
testing synthesis models for the X-ray background}

\author[Comastri et al.]{A. Comastri$^1$, F. Fiore$^{2,3}$,  
C. Vignali$^{1,4}$, G. Matt$^5$, G.C. Perola$^5$, F. La Franca$^5$ \\
$^1$ Osservatorio Astronomico di Bologna, via Ranzani 1, I--40127
Bologna, Italy \\
$^2$ Osservatorio Astronomico di Roma, Via Frascati 33,
I--00044 Monteporzio, Italy\\
$^3$ BeppoSAX Science Data Center, Via Corcolle 19, I--00131 Roma, Italy\\
$^4$ Dipartimento di Astronomia, Universit\`a di Bologna, via Ranzani 1, 
I--40127 Bologna, Italy\\ 
$^5$ Dipartimento di Fisica, Universit\`a degli Studi ``Roma Tre",
Via della Vasca Navale 84, I--00146 Roma, Italy \\
}

\maketitle
\begin{abstract}
The BeppoSAX High Energy Large Area Survey (HELLAS) has surveyed 
several tens of square degrees of the sky in the 5--10 keV band 
down to a flux of about $5\times10^{-14}$ \cgs.  
The source surface density of 16.9$\pm$6.4
deg$^{-2}$ at the survey limit corresponds to a resolved fraction of
the 5--10 keV X--ray background (XRB) of the order of 20--30 \%.
The extrapolation of the HELLAS logN--logS towards fainter fluxes
with an euclidean slope is consistent with the first XMM--{\it Newton} 
measurements, in the same energy band, which are a factor 20 
more sensitive.
The source counts in the hardest band so far surveyed by
X--ray satellites are used to constrain XRB models. 
It is shown that in order to reproduce the 5--10 keV 
counts over the range of fluxes covered by 
BeppoSAX and XMM--{\it Newton} a large fraction of 
highly absorbed (log$N_H$ = 23--24 cm$^{-2}$), 
luminous ($L_X > 10^{44}$ erg s$^{-1}$)
AGN is needed. A sizeable number of more heavily obscured, Compton thick,
objects cannot be ruled out but it is not required by the present data.
The model predicts an absorption distribution consistent with that 
found from the hardness ratios analysis of the so far identified 
HELLAS sources. 
Interestingly enough, there is evidence of a decoupling between 
X--ray absorption and optical reddening indicators especially 
at high redshifts/luminosities where several broad line quasars
show hardness ratios typical of absorbed power law models 
with log$N_H$=22--24 cm$^{-2}$.

\end{abstract}
 
 
\section{Introduction}

Hard X-ray surveys can permit to resolve directly the sources making
the hard Cosmic X-ray background (XRB) and so provide strong
constraints on unification schemes and AGN synthesis models for the
XRB.  According to these models a large fraction of the 2-100 keV
XRB energy density is due to obscured AGN (Setti \& Woltjer 1989;
Madau, Ghisellini \& Fabian 1994; Comastri et al. 1995; 
Gilli, Risaliti \& Salvati 1999; Wilman \& Fabian 1999; 
Pompilio, La Franca \& Matt 2000),
emerging at high energies.  A major step forward in the study
of the hard X--ray sky has been achieved by ASCA (Georgantopoulos et
al. 1997; Cagnoni Della Ceca, Maccacaro 1998; Boyle et al. 1998; Ueda
et al. 1998,1999; Della Ceca et al. 1999) and BeppoSAX (Giommi Perri \& 
Fiore 2000).  At the flux limit of the
deepest ASCA and BeppoSAX surveys in the 2--10 keV band
($\sim$ 5 $\times$ 10$^{-14}$ \cgs) about 30 \% of
the XRB is resolved into single sources. {\it Chandra} and 
XMM--{\it Newton} pushed the
2--10 keV flux limit $\approx20$ times fainter, resolving between 70
and 100 \% of the hard X--ray Cosmic background (Mushotzky et al. 2000,
Giacconi et al. 2001; Hasinger et al. 2001).  
Optical follow--up observations (Akiyama et
al. 2000; Fiore et al. 1999, 2000; Mushotzky et al. 2000; Giacconi et
al. 2001), revelead that a large fraction of these sources are AGN and
that a sizeable fraction of them show evidence of X--ray and optical
obscuration. Absorbed AGN can be detected, of course less
efficiently, also below 2 keV. Indeed, the optical identification
process of deep and ultra--deep ROSAT PSPC and HRI surveys (resolving
$\sim 70 \%$ of the 0.5--2.0 keV XRB, Hasinger et al. 1998), finds
that about one fourth of the sources are intermediate AGN or narrow
line galaxies, likely sites of high obscuration (the remaining sources
being broad line quasars, Schmidt et al. 1998, Lehmann et al. 2000).

In order to investigate the nature of the sources of the hard XRB
we have carried out a systematic search of X--ray sources in the
5--10 keV energy range, the hardest band accessible with imaging 
instruments. The High Energy Large Area Survey (HELLAS) is described 
in detail by Fiore et al. (2001, paper II). The hard X--ray selected 
sources have been extensively observed also at other wavelengths.
The results of optical identifications are reported by 
Fiore et al. (1999, paper I) and La Franca et al. (in preparation, paper V);
the search for soft X--ray counterparts in the ROSAT PSPC archive 
is reported by Vignali et al. (2001, paper IV), while the results 
of VLA radio observations are discussed by Ciliegi et al. (in preparation, 
paper VI).   
Here we show that the BeppoSAX and XMM--{\it Newton}
source counts in the 5--10 keV 
energy range and the average spectral properties of HELLAS sources 
provide further constraints on the absorption and luminosity distribution
of the sources making the hard XRB and on AGN synthesis models. 
We also show the need of a large number of highly
obscured (log$N_H$ = 23--24 cm$^{-2}$), high luminosity sources to reproduce 
the XRB spectrum and the number counts.
$H_0$ = 50 km s$^{-1}$ Mpc$^{-1}$ and $q_0$ = 0 are adopted throughout this 
paper.

\section{The BeppoSAX HELLAS survey}

The survey is presented in the companion paper by Fiore et al. (paper II). 
Briefly, the High
Energy Large Area Survey (HELLAS) covers about 85 square degrees in
the 5--10 keV band down to a flux limit of $\sim5\times10^{-14}$ \cgs.
Sources were detected and characterized statistically using the
methods described in the companion paper.  The quality of the
detection has always been checked interactively.  The count rates were
converted to fluxes using a fixed conversion factor equal to
7.8$\times10^{-11}$ \cgs (5--10 keV flux) per one ``3 MECS count''
(4.5--10 keV). Due to the narrowness of the band this factor is not
strongly sensitive to the actual spectral shape.  The sample used for
computing the logN-logS includes 147 sources. 
Optical identifications are available for about one third of the 
sample (La Franca et al. 2000).

\section{The XRB model}

The increasing amount of observational data concerning both 
the XRB spectral intensity and the source counts in several 
energy ranges obtained by ROSAT, ASCA and BeppoSAX surveys 
coupled with the optical identification of sizeable samples of 
hard X--ray selected sources (Fiore et al. 1999, 2000, Akiyama et al. 2000)  
have revived the interest on AGN synthesis models 
for the XRB  (Gilli et al. 1999; Wilman \& Fabian 1999; Pompilio et al. 2000).
A detailed treatement of XRB synthesis models
using the most recent observational constraints, 
including the AGN luminosity function derived from ROSAT 
deep surveys (Miyaji et al. 2000) and the column density distribution 
of Seyfert 2 galaxies in the local Universe estimated 
by Risaliti et al. (1999), is extensively discussed 
by Gilli et al. (2001), while a comparison between the different 
approaches is reviewed by Comastri (2000).
Even though a large fraction of the XRB spectral intensity 
and the source counts are fairly well reproduced by AGN synthesis models, 
a self--consistent description of the XRB constituents 
has yet to be reached as most of its ingredients have to be extrapolated 
well beyond the present observational limits. 

In almost all the AGN synthesis models for the XRB it is assumed 
that the space density and evolution of the unabsorbed population can 
be modeled by the QSO X--ray luminosity function computed from ROSAT 
surveys (Boyle et al. 1993,1994; Miyaji et al. 2000).
Obscured sources are then added with a distribution of column densities
and with the further assumption that their luminosity function and evolution 
are the same of unobscured population. 
Even though this approach seems reasonable, it is however plausible 
that absorbed objects could slip into deep ROSAT surveys at faint fluxes 
and thus skew the interpretation of the luminosity function and the 
modelling of the XRB.
The possible ``contamination'' of absorbed objects 
is minimized using the luminosity function parameters obtained from 
relatively shallow ROSAT surveys above $\sim$ 10$^{-14}$ erg cm$^{-2}$ 
s$^{-1}$ in the 0.5--2 keV band (Boyle et al. 1993, 1994)
 which are essentially dominated by unabsorbed objects (see also Fig.~4a).
The lack of a trend for a change in hardness ratio with redshift 
(Almaini et al. 1996) adds further weight to the assumption that this is 
a fair ``unobscured'' sample.

For the purpose of the present paper we adopt the Comastri et al. (1995
hereinafter C95) model with a few variations and improvements
and test its prediction against the HELLAS and XMM {\it Newton} 
survey findings.
With respect to C95 the spectrum of heavily absorbed 
AGN (log $N_H$ $>$ 24) is now computed taking into account Compton 
scattering effects by means of a Montecarlo code 
(Matt, Pompilio \& La Franca 1999).
The luminosity evolution, parameterized with a power law model 
$L(z) = L(z=0) \times (1 + z)^{2.6}$, stops at  
$z_{cut}$ = 1.8 and then stays constant up to $z_{max}$=3.
The shape and normalization of the luminosity function
are slightly different from those adopted in C95 and consistent 
with the values of Boyle et al. (1994).
The choice of these parameters is consistent with the 
more recent estimates of the high redshift AGN evolution
and also fits well the first {\it Chandra} measurements.
A more detailed comparison with observational data will be 
discussed in the following section(s).

The input AGN spectra, including the subdivision in the $N_H$ classes 
centered at log$N_H$ = 21.5,22.5,23.5,24.5, are as described in   
C95, while the best fit absorption distribution of obscured AGN 
in the four $N_H$ classes normalized to the space density of unobscured AGN, 
turned out to be 0.35,1.5,2.3,2.0 which is slightly different from that 
reported by C95 for two reasons. The first is that 
some input parameters have been changed 
as described in the previous paragraph, 
the second concerns the intensity of the 1--8 keV XRB spectrum 
which has to be fitted by the model. 
The ASCA (Gendreau et al. 1995) and BeppoSAX (Vecchi et al. 1999) 
observations of the XRB spectrum below
8 keV are significantly higher than the 
average level measured by HEAO1--A2 (Marshall et al. 1980). 
The maximum discrepancy is between the BeppoSAX and HEAO1--A2 
observations, the level of the latter being lower by 30--40 \% in 
the overlapping 3--8 keV band, The best fit ASCA spectrum lies approximately
in between and thus we choose to fit the XRB spectral intensity 
as measured by ASCA. 

\begin{figure}
\centerline{
\psfig{file=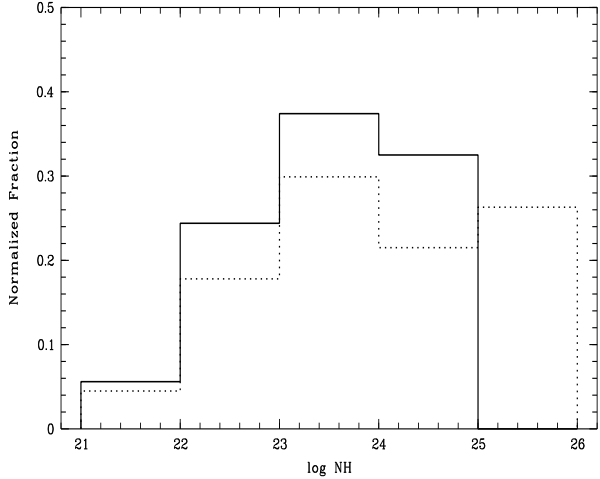, width=8cm, height=9cm, angle=-90}
}
\caption{The model absorption distribution (solid line) 
compared with the observed $N_H$ distribution for a sample of
nearby Seyfert 2 galaxies (Risaliti et al. 1999). The two histogram are 
renormalized to the same ratio between obscured and unobscured objects.}
\label{abs_comp}
\end{figure}

The best fit absorption distribution is compared with that derived 
by Risaliti et al. (1999) in Fig.~\ref{abs_comp}. 
Even though the relative fraction of 
objects in the various $N_H$ bins is different the overall
shape of the two distributions is similar.
The peak in both cases is in the log$N_H$ = 23--24 bin and a 
large fraction of the objects (70--80 \%) is obscured by column densities 
greater than 10$^{23}$ cm$^{-2}$.
As a consequence the large fraction of highly obscured objects 
needed to fit the XRB spectral intensity is consistent with 
the absorption distribution of Seyfert 2 galaxies in the local Universe.

Sources with column densities in 
excess of 10$^{25}$ cm$^{-2}$ are not included in the present model
as their contribution to the XRB is never energetically relevant  
(see e.g. Gilli et al. 1999) even if they could constitute a sizeable 
fraction of the obscured AGN population (Risaliti et al. 1999).

\section {The integral logN-logS compared with model predictions}

\begin{figure}
\centerline{
\psfig{file=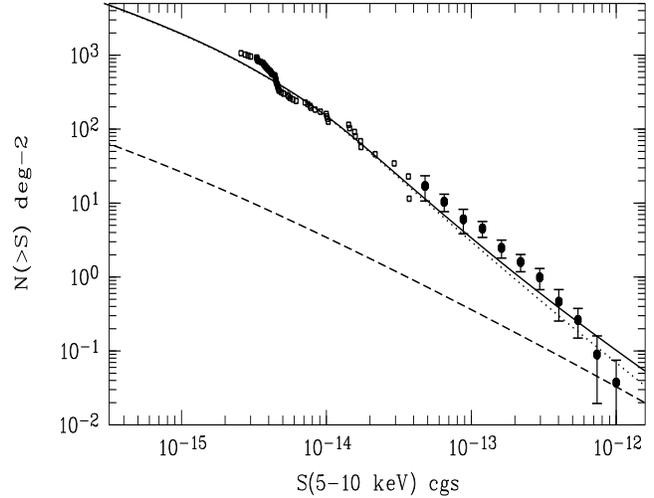, width=8cm, height=9cm, angle=-90}
}
\caption{The 5--10 keV integral logN-logS as measured by BeppoSAX 
(filled circles with error bars) and XMM--{\it Newton} (open squares).  
Thick line = total (AGN+clusters) 5--10 keV counts predicted by 
the XRB synthesis model; dotted line = AGN only; dashed line = 
clusters only.
}
\label{lnls510}
\end{figure}

Figure \ref{lnls510} presents the integral 5--10 keV logN-logS of the 147
HELLAS sources. The
logN-logS can be represented by a power law model 
$N(>S) = K S^{-q}$ below a flux of 
$4\times10^{-13}$ \cgs. The best fit power law index is
$q=1.56\pm0.14(0.34)$. Errors represent the 90 \% confidence
interval, errors in brackets include systematic uncertainties.  The
normalization $K$ of the best fit power law at 10$^{-13}$ erg cm$^{-2}$
s$^{-1}$ is 5.24 deg$^{-2}$.
Above $\sim4\times10^{-13}$ \cgs the
logN-logS steepens significantly. This is probably due to a selection
effect. The probability to find a bright source near a, usually,
bright target is small, and therefore the number of bright sources in
a serendipity survey like the HELLAS survey tends to be
underestimated. 
The derived shape of the logN-logS may be affected by several
biases which are extensively discussed in paper II. 
For the purposes of the present discussion
we compare the model predictions with the best estimate of 
HELLAS counts taking into account both statistical and systematic errors.    

\begin{figure}
\centerline{
\psfig{file=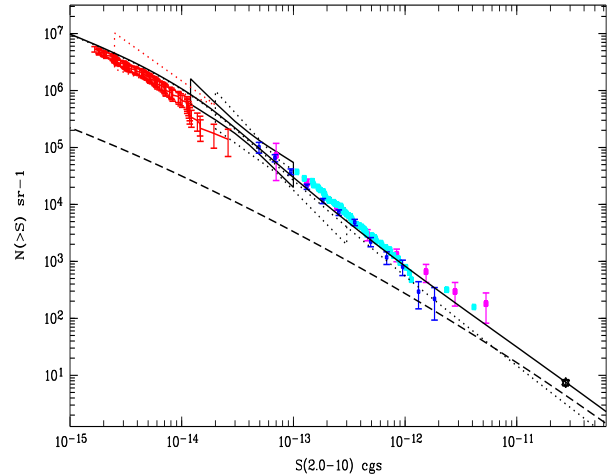, width=8cm, height=9cm, angle=-90}
}
\caption{A compilation of the 2--10 keV logN-logS from different 
surveys. Blue dots = BeppoSAX, Giommi et al. 2000; 
magenta squares = ASCA, Ueda et al. 1999; cyan squares = ASCA, Della
Ceca et al. 1999; red error bars = Chandra, Giacconi et al. 2001;
black star = HEAO1, Piccinotti et al. 1982;
black thick bow tie = the 1$\sigma$ region delimited by the analysis of
BeppoSAX fluctuations, Perri \& Giommi 2000;
black dot bow tie = the 1$\sigma$ region delimited by the 
analysis of ASCA SIS fluctuations, Gendreau et al. 1998;
red dot bow tie = Chandra logN-logS using eq. 2 of Mushotzky et
al. 2000).
Thick line = total (AGN+clusters) 2-10 keV counts predicted by 
the XRB synthesis model; dotted line = AGN only; dashed line = 
clusters only.
}
\label{lnls210}
\end{figure}

The solid thick line in fig. \ref{lnls510} represents the
expectation of the model described in the previous section. This
includes both AGN and clusters. The expected number counts relation for
cluster of galaxies has been computed assuming an average temperature
of 6 keV and the local luminosity function of Piccinotti et al. (1982)
with no evolution. The cluster contribution, which is never dominant 
at the 5--10 keV fluxes sampled by BeppoSAX, is reported only 
for completeness. We note that a detailed calculation of cluster
source counts using more recent estimates for the clusters XLF 
(Ebeling et al. 1997) provides similar results.  
The predicted hard X--ray counts are also consistent with the 
recent XMM--{\it Newton} observations in the Lockman hole 
reaching 5--10 keV fluxes about an order of magnitude fainter
(Hasinger et al. 2001). 
The expectation of the same model are also compared with a compilation of 
number counts in the 2--10 keV band (Fig. \ref{lnls210})
from BeppoSAX (Giommi et al. 2000), ASCA (Ueda et al. 1999, 
Della Ceca et al. 1999) and Chandra (Mushotzky et al. 2000, 
Giacconi et al. 2001) surveys. A good agreement between model predictions 
and observations 
is obtained with the BeppoSAX and ASCA  2--10 keV number counts, 
and with the Mushotzky et al. (2000) {\it Chandra} counts while 
the measurements of Giacconi et al. (2001) are slightly overpredicted.

\begin{figure}
\centering{
\psfig{file=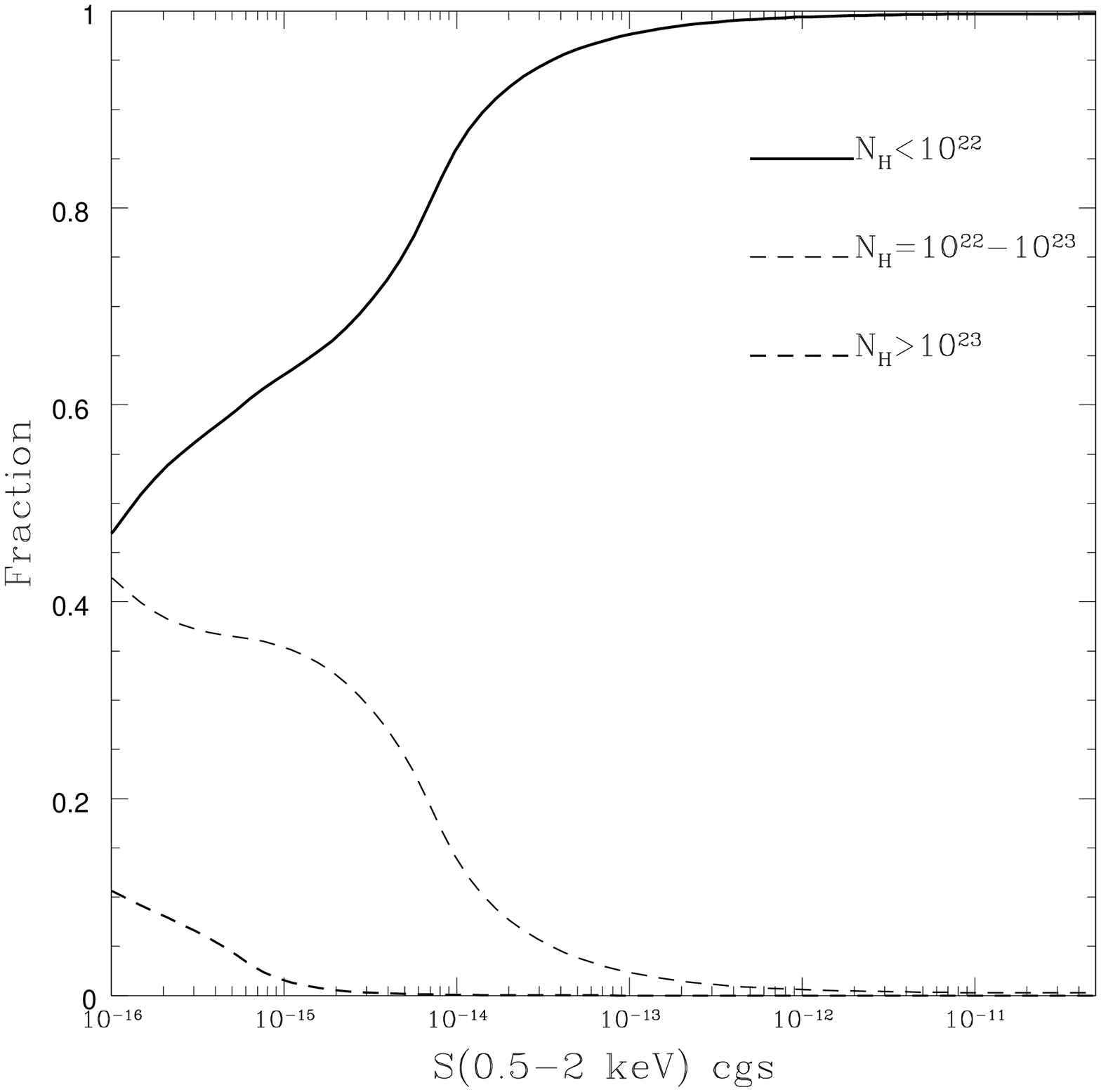, width=0.4\textwidth, angle=0}
\psfig{file=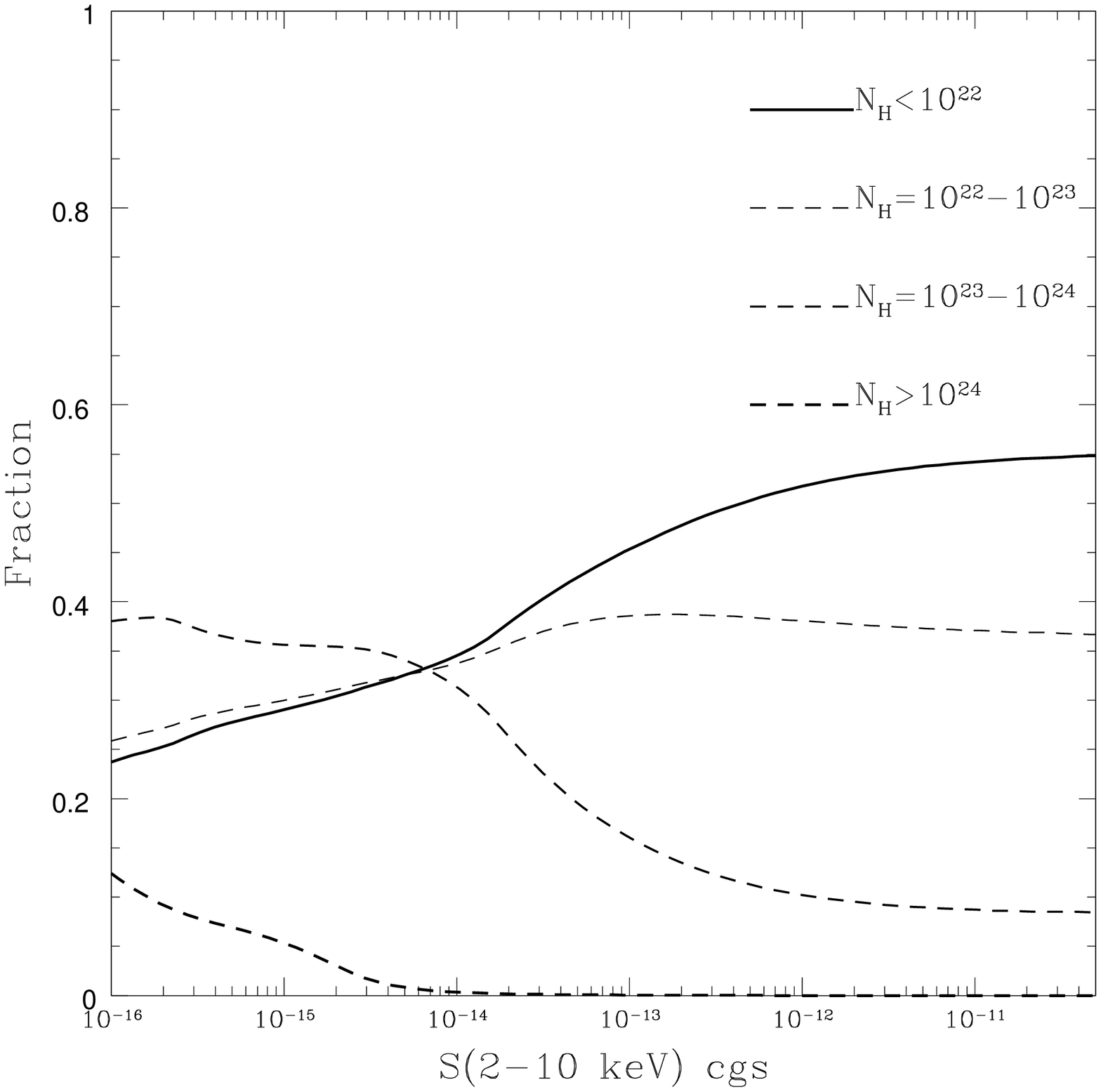, width=0.4\textwidth, angle=0}
\psfig{file=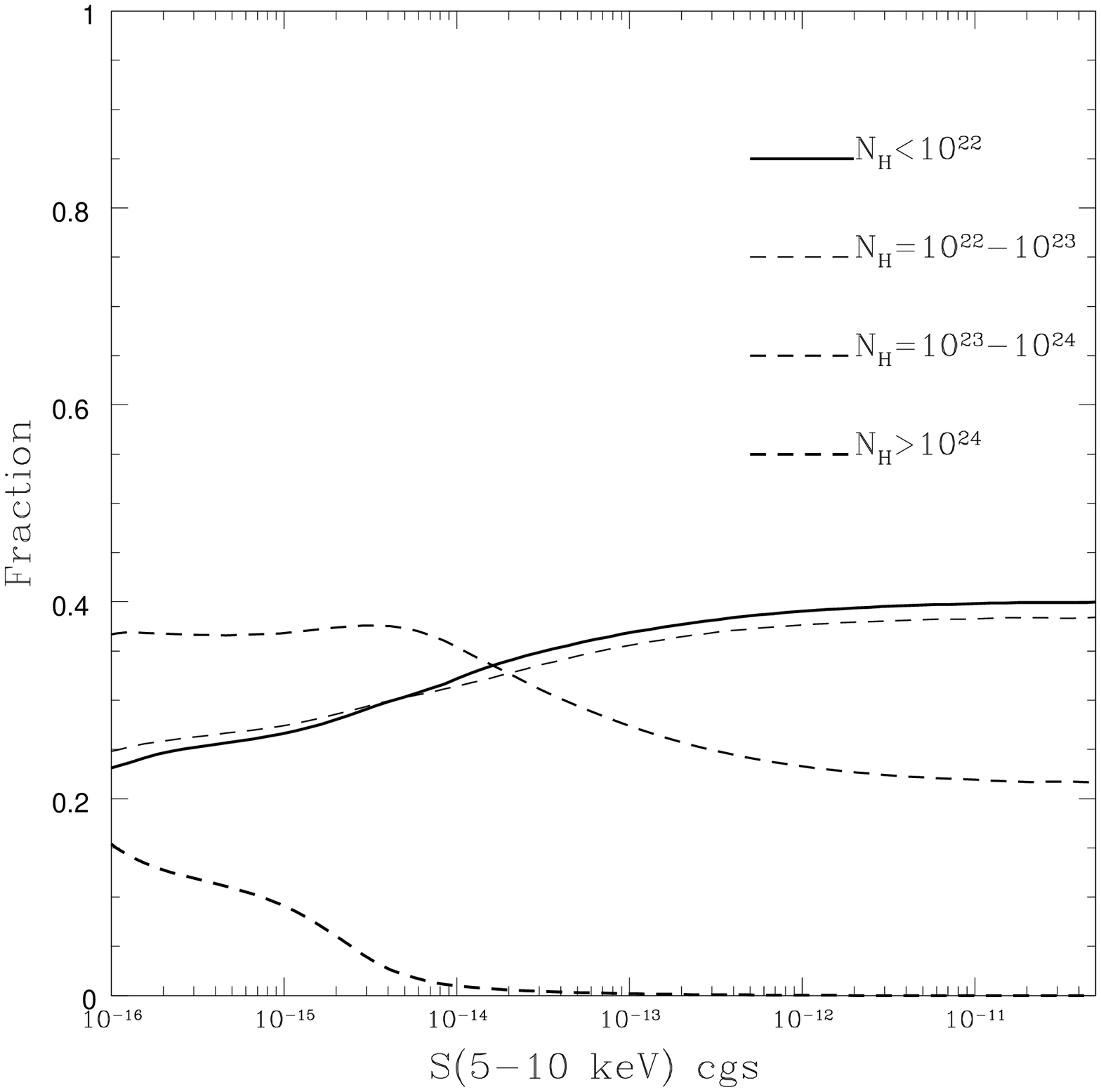, width=0.4\textwidth, angle=0}
}
\caption{ The relative fraction of unobscured (solid line) and
obscured sources (dashed lines where the thickness increases for
increasing absorption) as a function of the 0.5--2, 2--10 and
5--10 keV flux from top to bottom respectively.}
\label{fraction}
\end{figure}

The key parameter of all the XRB models is the space density of
obscured AGN and their absorption distribution.
It is customary to adopt a ratio between 
obscured and unobscured objects consistent with the 
value measured for samples of nearby (often optically selected) 
type 2 and type 1 Seyferts galaxies.
In the model discussed here such a ratio is not assumed a priori
but determined from the XRB fitting procedure 
and thus it is obviously dependent on the column density adopted 
to divide absorbed from unabsorbed objects.
The ratio between obscured and unobscured AGN is thus about 4.3 
if such a value is set to $10^{22}$ cm$^{-2}$ and about 2.4 
for $N_H$ = 3 $\times$ 10$^{22}$ cm$^{-2}$.
We stress that such a ratio is independent from optical classification 
and thus the relative fraction of X--ray obscured AGN in a flux limited 
survey could be different from the fraction of optically narrow--lined 
type 2 objects. It is important also to note that the above quoted 
values for the ratio between obscured and unobscured objects 
are referred to the entire AGN population responsible for the 
XRB intensity and that such a number would be recovered only 
when the full XRB is resolved into single sources.
The fraction of objects
with a given column density actually included in a survey 
depends on the flux limit of the survey and on the energy range. 
Figure \ref{fraction} shows these fractions as a function of the
0.5--2, 2--10 and 5--10 keV fluxes. 
The dependence of the fractions on the flux 
is smoother in the harder bands (see Comastri 2000).
As a consequence, 5-10 keV surveys can better probe 
the real fraction of obscured object going down to fluxes
comparatively higher than in the 2-10 keV and 0.5-2 keV bands.  
As an example at 5--10 keV fluxes of the order of 10$^{-13}$ erg cm$^{-2}$
s$^{-1}$   
the fraction of objects in the log$N_H$ = 23--24 bin is of the order
of 25\% while is some $<$ 15 \% in the 2--10 keV band and negligible
in the soft 0.5--2 keV energy range.

\subsection{The luminosity and absorption distribution of obscured AGN} 

In order to investigate to what extent the present data 
can be used to constrain the absorption and luminosity distribution 
of the sources making the hard XRB,
additional tests of synthesis models have been carried out.
More specifically we have computed 2 models which 
both reproduce the XRB spectral shape over the 3--100 keV energy range 
with somehow ``extreme'' assumptions on the absorption and 
luminosity distribution of the obscured sources.
In the former case we have maximized the number of  
obscured Compton thick sources (about a factor 2 with respect to the 
baseline model), while in the latter 
a cut--off in the luminosity function of absorbed AGN 
has been introduced for luminosities greater than 10$^{44}$ erg s$^{-1}$.
These attempts are motivated by both theoretical and observational 
evidence as described below.  

An XRB model dominated by a population of almost Compton thick
($\tau_T \simeq$ 1) sources has been worked out by Wilman et al. (2000)
based on the suggestion of Fabian (1999) who argued that 
significant obscuration might be associated 
with the growth of massive black holes in proto--galactic bulges.
The Wilman et al. (2000) model predicts a similar contribution
from obscured and unobscured objects at about 30 keV.
In order to reproduce the XRB spectrum below about 10 keV additional 
obscuration with an average column of $N_H$ = 10$^{22.75}$ cm$^{-2}$ 
is introduced for 70\% of the remaining objects.

We have computed a synthesis model with an absorption distribution 
dominated by Compton thick sources. 
The contribution of these sources to the XRB is normalized 
to that obtained by Wilman et al. (2000, see their figure 1)
while the absorption distribution for sources with lower column densities 
is tuned to keep the overall number of objects in the various $N_H$ 
classes equal to that of our baseline model and to fit the XRB spectrum.  

\begin{figure}
\centerline{
\psfig{file=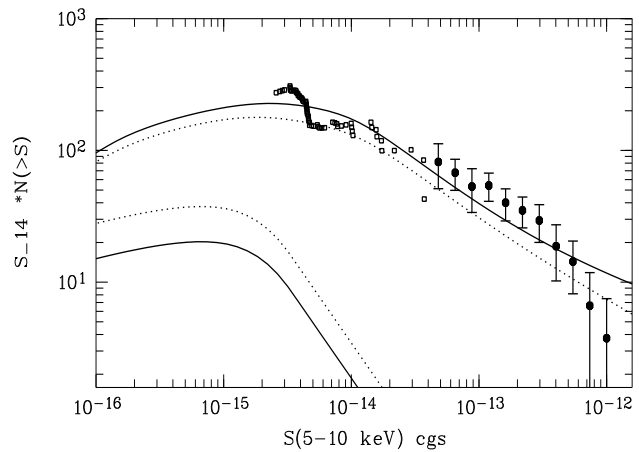, width=7cm, height=9cm, angle=-90}
}
\caption{The 5--10 keV  HELLAS and XMM--{\it Newton} counts are compared
with the Compton--thick--dominated model (dotted lines) 
and with the baseline model (solid lines)
predictions. The contribution of  Compton thick sources 
in the two models is also reported.}
\label{sns}
\end{figure}

The 5--10 keV predicted counts are reported in 
Figure \ref{sns} and compared with the observed HELLAS counts and with the
baseline model predictions. The y--scale is the product of the 
integral counts times the flux. In this 
representation the fluxes contributing most to the XRB 
are around the peak of the distribution.

The contribution of Compton thick sources is never dominant 
in the 5--10 keV band, as a consequence, the bright part of the 
logN--logS is not well reproduced by the model where their contribution
is maximized. Even though the uncertainties associated with the
hard HELLAS and XMM--{\it Newton} counts do not allow to rule out 
a large number of highly obscured AGN, it is clear that 
on the basis of present data an $N_H$ distribution peaked 
at lower column densities has to be preferred.

\begin{figure}
\centerline{
\psfig{file=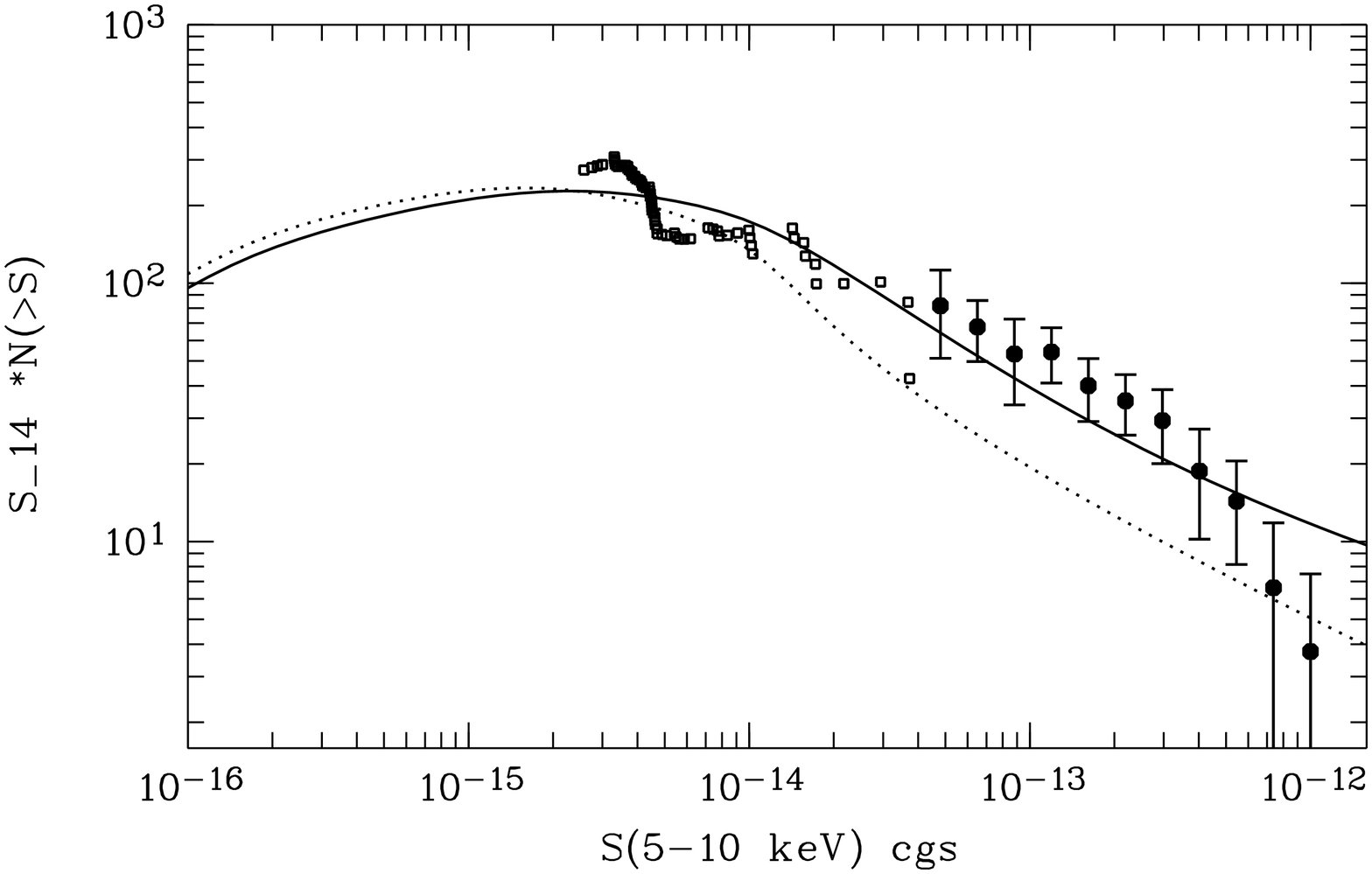, width=8cm, height=9cm, angle=-90}
}
\caption{
The 5--10 keV  HELLAS and XMM--{\it Newton} counts are compared
with the predictions of a model without high luminosity highly obscured
objects (dotted line) and with the baseline model.}
\label{sns_l44}
\end{figure}

The evolution of the luminosity function of absorbed sources 
is assumed to be the same of unobscured one in almost all the 
XRB synthesis models. The lack of luminous narrow line quasar (type 2 
QSO) in optical surveys has lead several authors (see the discussion 
in Halpern et al. 1998) to conclude that either they are very rare 
or even do not exist. We checked the importance of high--luminosity 
highly absorbed AGN by fitting the XRB with a mixture of 
unabsorbed AGN at all luminosities and low--luminosity 
absorbed AGN with an approach similar to that described by 
Gilli et al. (2001).
Here a sharp cut--off in the luminosity function of obscured AGN
is introduced for luminosities greater than 10$^{44}$ erg s$^{-1}$.
The shape of the absorption distribution is the same of the baseline model; 
however, in order to reproduce the XRB spectral intensity, the number 
of sources in each absorption class has to be increased by 20\%.
The predicted 5--10 keV counts are reported in Fig.~ \ref{sns_l44}
and compared with the observational data and with the baseline model.
It is clear that with such a model the HELLAS counts at relatively 
bright fluxes are severely underestimated since most of the XRB 
energy density is due to low luminosity AGN emerging only at 
lower fluxes (Comastri 2000).

\section{The HELLAS average spectral properties compared with 
model predictions}

\begin{figure*}
\centerline{
\psfig{file=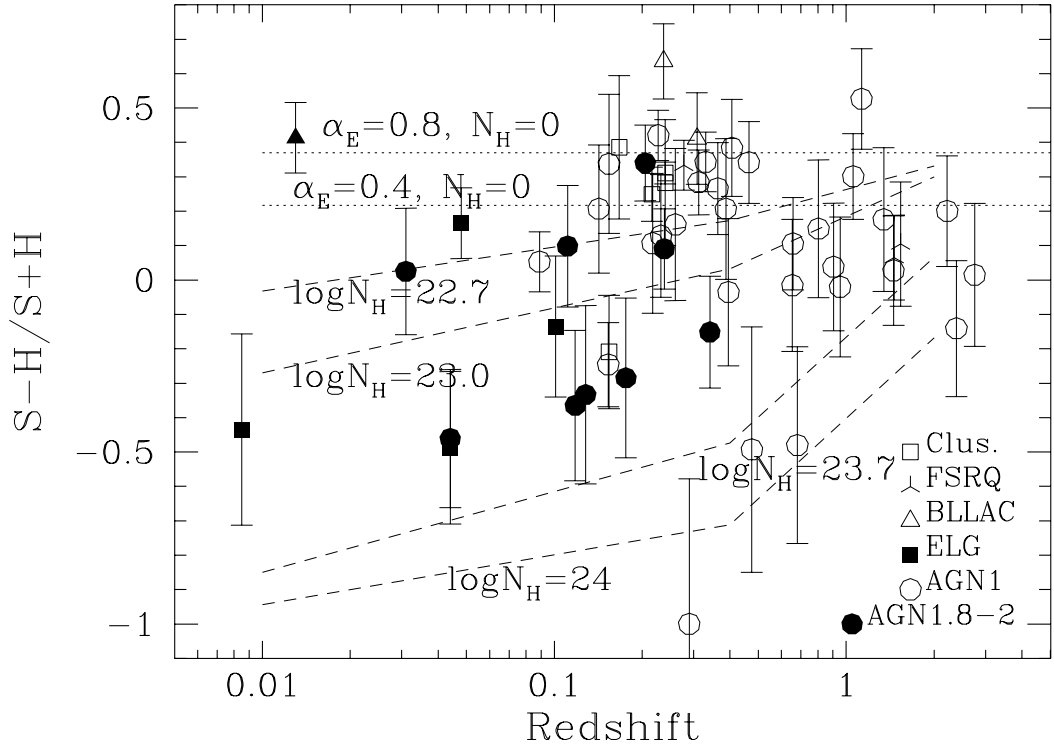, width=0.7\textwidth, angle=-90}
}
\caption{ The softness ratio (S-H)/(S+H) as a function of redshift for
the identified sources. Different symbols mark identified sources:
open circles = broad line, `blue' continuum quasars and Sy1; stars=
broad line `red' continuum quasars; filled circles= type 1.8-1.9-2.0
AGN; filled squares = starburst galaxies and LINERS; 
open triangles = radio loud AGN. The dotted lines show
the expected hardness ratio for power law models with $\alpha_E$=0.8 and 0.4.
Dashed lines show the expectation of absorbed power law models 
(with $\alpha_E=0.8$ and log$N_H$=24, 23.7, 23.0, 22.7 from bottom to top) 
with the absorber at the source redshift.}
\label{hrtz}
\end{figure*}

To evaluate the absorbing column density of the HELLAS sources from
the X-ray count ratio, the redshift of the source should be known
(assuming that the absorber is at the same redshift of the source, see
Fiore et al. 1998). To this purpose we plot (S$-$H)/(S+H) as a function
of redshift (Figure \ref{hrtz}) for a subsample of 51 sources with
optical spectroscopic identification (La Franca et al. 2001).
The dotted lines represent the expectation of unabsorbed
power laws with $\alpha_E=0.8$ and 0.4.  The dashed lines represent the
expectations of power law absorbed by columns of  
$10^{23}$, $10^{23.5}$ and $10^{24}$ cm$^{-2}$, respectively, 
{\it in the source frame}.  
As expected, the softness ratio of constant column
density models strongly increases with the redshift. Many of the
identified HELLAS sources have (S--H)/(S+H) inconsistent with that
expected from an unabsorbed power law model with $\alpha_E=0.8$. 
Substantial absorbing column densities are implied.

Not surprisingly, almost all the sources in the nearby 
Universe ($z <$ 0.3), which are likely to be obscured by
column densities $N_H > 10^{22.5-23}$ cm$^{-2}$, 
are narrow line, type 1.8-2 AGN 
(Fiore et al. 1999, 2000; La Franca et al. 2001), while 
broad line AGN are relatively unabsorbed in agreement with 
popular AGN unification schemes. 
The situation is different at high redshift where several 
broad--line AGN appear to have strongly absorbed X--ray spectra.
The statistics in the present observations
are not good enough to definitely assess this point. XMM--{\it Newton} 
observations may easily confirm or disprove a significant 
absorbing column in these sources.
The presence of highly X--ray obscured quasars with broad optical lines
implies a dust--to--gas ratio or a dust composition strongly different 
from the Galactic one (Maiolino et al. 2001). Alternatively,  
the X--ray absorber could be much closer to the central source than the
Broad Line Region within the dust sublimation radius. This radius
depends on the central source luminosity and thus if the 
X--ray absorbing gas lies at the same distance from the ionizing 
source, the higher the X--ray luminosity the lower the 
dust extinction is. Although the nature of X--ray absorption 
in broad lines blue continuum quasars is not well understood, 
it is interesting to note that large column of cold gas have been detected
in broad absorption line quasars (Gallagher et al. 1999), bright PG 
quasars (Gallagher et al. 2000) and high redshift quasars 
(Elvis et al. 1994; Cappi et al. 1997; Fiore et al. 1998; 
Yuan et al. 2000; Fabian et al. 2000).

\begin{figure}
\centerline{
\psfig{file=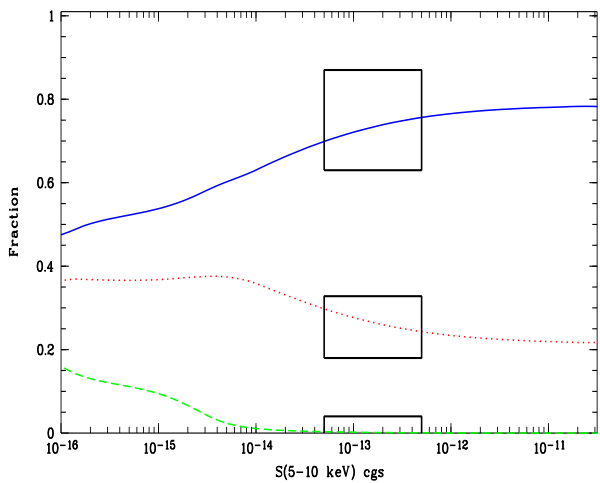, width=8cm, height=9cm, angle=-90}
}
\caption{ The model predicted fraction of relatively unobscured 
(log $N_H < $ 23, solid line), highly absorbed (23 $<$ log $N_H$ $<$ 24, 
dotted line) and Compton Thick 
(log $N_H > $ 24, dashed line) AGN as a function of the 5--10 keV flux. 
The boxes corresponds to the fraction of objects in the
corresponding $N_H$ classes as computed from the softness ratio vs. redshift
diagram of figure 7.
}
\label{abs_pred}
\end{figure}

The column density distribution inferred from Fig.~\ref{hrtz} is 
compared with 
the baseline model predictions in Fig~\ref{abs_pred}. Given that 
the 5--10 keV band is not sensitive to column densities below
about $10^{23}$ cm$^{-2}$, sources with lower column densities
are grouped in a single bin. The two BL Lac objects have been 
neglected in the calculation as this class of objects 
is not included in the synthesis model.  
The two boxes correspond to the observed fraction ($\pm$1 $\sigma$) 
of HELLAS sources for each absorption class and cover the flux decade
$5\times10^{-14}$--$5\times10^{-13}$ corresponding to the fluxes 
of the large majority of the sources in Fig.~\ref{hrtz}.
In order to take into account the relatively large errors in the 
softness ratio values    
a source is included in the corresponding $N_H$ class only if 
its softness ratio value plus one sigma error is below the 
$N_H$ value indicated by the dashed lines in Fig~\ref{hrtz}. 
According to this prescription there are no sources with 
column densities greater than $10^{24}$ cm$^{-2}$, the corresponding 
upper limit was obtained assuming that the two objects 
with a ``best fit'' column density $>10^{24}$ cm$^{-2}$
are indeed Compton thick. It is important to note that 
the relatively good agreement between model predictions 
and observed $N_H$ distribution is obtained without considering the 
optical classification.

\section{Discussion and Conclusions}

The HELLAS survey described in the companion paper by Fiore et al. (paper II) 
provided a statistically well
defined and flux limited sample of 147 hard X--ray selected sources 
used to estimate the 5--10 keV logN--logS and the average 
X--ray spectral properties  
of the objects responsible for a significant fraction 
of the 5--10 keV XRB (from 20 to 30\% depending from its normalization;  
Vecchi et al. 1999, Comastri 2000).
The analysis of X--ray colors indicates that HELLAS sources 
have rather hard spectra, harder than 
the average spectra of sources in the 0.7--10 keV ASCA surveys 
(Ueda et al. 1999, Della Ceca et al. 1999).  
This hardness may be due to substantial
absorbing columns.  In fact, a large fraction of identified type 1.8-2
AGN and also a few broad line Seyfert 1 galaxies and quasars
show softness ratios similar to those expected from power law
models reduced at low energy by column densities of log$N_H$=22--24,
at the source redshift.  

The relatively high number of hard sources discovered by BeppoSAX 
in the hardest band accessible with 
imaging instruments makes the HELLAS survey extremely well suited 
to test AGN synthesis models predictions. 

The baseline model described in detail in previous sections
is able to reproduce the 5--10 keV counts as observed by BeppoSAX 
and XMM--{\it Newton} and is also consistent with the 2--10 keV 
logN--logS as measured by several instruments.
The bright tail of the 5--10 keV logN--logS as measured by BeppoSAX  
is best fitted by a model with a significant fraction of highly 
obscured, high luminosity AGN in good agreement with the 
findings of Gilli et al. (2001) based on the 2--10 keV counts.
A large fraction of Compton thick sources, as envisaged in the model 
by Wilman et al. (2000) and Fabian (1999), does not adequately reproduce 
the HELLAS counts, however the presence of a sizeable number 
of sources with $N_H > 10^{24}$ cm$^{-2}$ cannot be strongly ruled out based
on the 5--10 keV counts. Imaging observations at energies greater 
than about 20 keV, such as those planned with 
the Energetic X--ray Imaging Survey Telescope (EXIST, Grindlay 2000),
are required to test such possibility.

The baseline model can be considered as the simplest 
version of AGN synthesis model. In fact the 
absorption distribution adopted to fit the hard 
XRB spectrum and the source counts in different energy ranges 
are independent from source redshift and luminosity. 
Moreover, such a distribution 
is also in good agreement with that derived from observations of nearby 
AGN (Risaliti et al. 1999). 
 
The optical identification of a sizeable number of HELLAS 
sources allowed to further constrain the absorption distribution, 
for column densities lower than $\simeq$ 10$^{24}$ cm$^{-2}$, 
over a large range of redshifts and luminosities.
The most important result concerns the optical appearence 
of X--ray obscured AGN as a function of redshift.
In the local Universe absorbed objects are almost always 
associated with narrow line optical spectra while 
at high redshifts ($z >$ 0.3) the presence of 
X--ray obscured sources with broad optical lines 
strongly suggest a decoupling between the 
X--ray and optical classification.
We note that such a result does not affect the 
AGN synthesis model which is entirely based on the 
X--ray properties.    
It is concluded that, as far as the present observational 
constraints are concerned, there is no need for a significant 
revision of the baseline model assumptions
which turn out to adequately reproduce both the XRB spectrum 
and the source counts. 

The optical identification of a sizeable sample of hard
X--ray selected sources from 
{\it Chandra} and XMM--{\it Newton} surveys coupled with high quality 
X--ray spectroscopy, especially for high redshift quasars,
will provide new insights on the nature and the physics of
the sources of the XRB.

\bigskip
\centerline{\bf Acknowledgements}

We thank the BeppoSAX SDC, SOC and OCC teams for the successful
operation of the satellite and preliminary data reduction and
screaning. We also thank R. Maiolino, S. Molendi, F. Pompilio,
M. Perri, M. Salvati and G. Zamorani for useful discussions
and G. Hasinger for providing the XMM--{\it Newton} counts 
in the 5--10 keV band in a computer readable format.
This research has been partially supported by ASI contract
ARS--99--75 and MURST grants Cofin--98--032 and Cofin00--02--36.

\end{document}